\def\year{2021}\relax
 
\documentclass[letterpaper]{article} 
\usepackage{aaai21}  
\usepackage{times}   
\usepackage{helvet}  
\usepackage{courier}  
\usepackage[hyphens]{url}  
\usepackage{graphicx} 
\usepackage{natbib}  
\usepackage{caption} 
\title{Fairness as Equality of Opportunity:\\ Normative Guidance from Political Philosophy}

\author {
    Falaah Arif Khan$^1$,
    Eleni Manis$^2$, and
    Julia Stoyanovich$^1$ \\
}
\affiliations {
    $^1$New York University\\
    $^2$Surveillance Technology Oversight Project \\
    fa2161@nyu.edu,
    em579@nyu.edu,
    stoyanovich@nyu.edu
}

\usepackage[utf8x]{inputenc}
\usepackage{xspace}
\usepackage{makecell}
\usepackage{xcolor}
\usepackage[position=bottom]{subfig}
\usepackage{float}
\usepackage{graphics}
\usepackage{graphicx}
\usepackage{color, colortbl}
\usepackage{caption}
\usepackage{wrapfig}
\usepackage{paralist}
\usepackage{booktabs} 
\usepackage{array}
\usepackage{balance}
\usepackage{lipsum}
\usepackage{optidef} 
\usepackage{dsfont} 
\makeatletter
\newcommand{\removelatexerror}{\let\@latex@error\@gobble}
\makeatother

\usepackage{amsmath}
\usepackage{mathtools}

\usepackage{tikz}
\usepackage{physics}
\usepackage{mathdots}
\usepackage{yhmath}
\usepackage{cancel}
\usepackage{siunitx}
\usepackage{array}
\usepackage{multirow}
\usepackage{amssymb}
\usepackage{tabularx}

\usetikzlibrary{arrows.meta}
\usetikzlibrary{positioning}
\usetikzlibrary{fadings}
\usetikzlibrary{patterns}
\usetikzlibrary{shadows.blur}
\usetikzlibrary{shapes}
\usetikzlibrary{shapes.symbols} 
\usepackage{booktabs, adjustbox}
\usepackage{color, colortbl}

\usepackage{bbm}

\newcommand{\adslong}{Automated Decision Systems\xspace}
\newcommand{\ads}{ADS\xspace}

\begin{document}

\maketitle

\begin{abstract}
Recent interest in codifying \emph{fairness} in \adslong (\ads) has resulted in a wide range of formulations of what it means for an algorithmic system to be \emph{fair}. Most of these propositions are inspired by, but inadequately grounded in,  political philosophy scholarship. This paper aims to correct that deficit. We introduce a taxonomy of fairness ideals using doctrines of Equality of Opportunity (EOP) from political philosophy, clarifying their conceptions in philosophy and the proposed codification in fair machine learning. We  arrange these fairness ideals onto an EOP spectrum, which serves as a useful frame to guide the design of a \emph{fair} ADS in a given context. 

We use our fairness-as-EOP framework  to re-interpret the impossibility results from a philosophical perspective, as the incompatibility between different value systems, and demonstrate the utility of the framework with several real-world and hypothetical examples. Through our EOP-framework we hope to answer  what  it means for an ADS to be \emph{fair} from a moral and political philosophy standpoint, and to pave the way for similar scholarship from ethics and legal experts. 

\end{abstract}
\section{Introduction}
\label{sec:intro}

\adslong (\ads) are broadly used socio-techno-political systems \cite{stoyanoj_mirror}, and codifying \emph{fairness} in the context of these systems requires a harmonization of scholarship in machine learning and political philosophy. While most propositions of \emph{fair machine learning} (fair-ML) draw on scholarship from political and moral philosophy, a critical survey of literature indicates both a naïve understanding of philosophical theories and a misunderstanding of their applicability to real-world contexts. The goal of this paper is to ground current (and future) approaches in fair-ML \cite{chouldechova_frontiers} in a better understanding of their counterparts in political philosophy, and to provide normative guidance as to which notion of fairness is applicable in which context, using the moral framework of Equality of Opportunity (EOP). While the ideas that follow are primarily theoretical, their implications are significant both for researchers and practitioners. 

Lifting the fog hanging over the fairness debate requires a willingness to engage in methodical moral reasoning. As observed by \citet{jacobswallach}, when the fairness debated is ``couched indirectly in mathematics,'' the moral nature of fairness is ``rendered less accessible to [nontechnical] stakeholders.''  We further argue that couching the fairness debate in mathematics hides the moral commitments underlying different fairness notions from fair-ML practitioners themselves. The EOP principle was first studied in fair-ML using economists' models of EOP \cite{heidari_EOP}.  We agree with the choice of principle but resist the reduction of moral reasoning to mathematical proofs because it obscures the moral views that practitioners must be able to articulate to make meaningful choices among fairness formulations. 

EOP is a philosophical principle that dictates how desirable positions (or ``opportunities") should be distributed in society. An EOP-respecting society eliminates barriers to achievement that are irrelevant and arbitrary, and different conceptions of EOP define different rules for a \emph{fair} competition. As a moral framework, EOP allows us to see, in an organized and comparative way, fairness notions’ motivations, strengths, and shortcomings.  It poses moral questions that fair-ML practitioners must take a stand on and guides practitioners toward fairness conceptions that satisfy their value judgments.  Faced with the impossibility results \cite{Kleinberg_impossibility, chouldechova_impossibility} for different statistical measures of fairness, we can make a principled choice, on the basis of well-articulated moral reasons, to drop the pursuit of a redundant fairness notion or to trade off between two competing fairness notions. The EOP framework also allows computer scientists to appreciate why their fairness-related choices may be rejected by stakeholders with different moral beliefs about what fairness and justice require.

\subsubsection{For fair-ML researchers}
The work of prominent political philosophers and economists such as Rawls, Roemer, Dworkin and Arneson continues to motivate new measures of \emph{fairness}. While many papers propose technical definitions inspired by philosophical doctrines, a typical paper substitutes mathematical proofs for moral reasoning once its fairness notion is operationalized. We introduce a taxonomy of fairness ideals using doctrines of EOP from political philosophy, clarifying their conceptions in philosophy and the proposed codification in fair-ML. 

\subsubsection{For practitioners}
Given the variety and criticality of contexts in which ADS are being used today, it is imperative to be able to identify the most suitable fairness criterion for a given application, and to have intuitive understanding of the values and moral principles that it encodes. The EOP-framework we propose helps guide the choice of fairness metrics for specific contexts. We hope that this will aid practitioners' efforts to connect technical definitions of fairness to the values that motivate them and thus aid efforts to communicate with stakeholders outside of computer science (such as lawmakers and policy makers).\par

\subsubsection{Paper organization} We begin by motivating why Equality of Opportunity (EOP) doctrines are a useful lens for organizing fairness concerns. We go on to introduce different doctrines of EOP, clarifying their conception (principles, intuitive appeal and challenges) in political philosophy, and contrasting it with the proposed codification in fair-ML. This allows us to identify and organize justice considerations that the fair-ML literature currently misspecifies, overlooks or underemphasizes. Next, we use this taxonomy to create an EOP Spectrum on which practical fairness concerns lie. We use the spectrum to reinterpret the recent impossibility results from a philosophical perspective. We then apply our EOP framework to a range of hypothetical and real world scenarios, demonstrating how EOP guides the selection of a suitable fairness ideal for a given context. We conclude with a look at the broader justice concerns, as encoded by Rawls's theory of justice, and with a summary of limitations of the fairness-as-EOP framework.
\section{Equality of Opportunity Doctrines}
\label{sec:eop}
\begin{figure}[ht]
\centering
\includegraphics[scale=0.42]{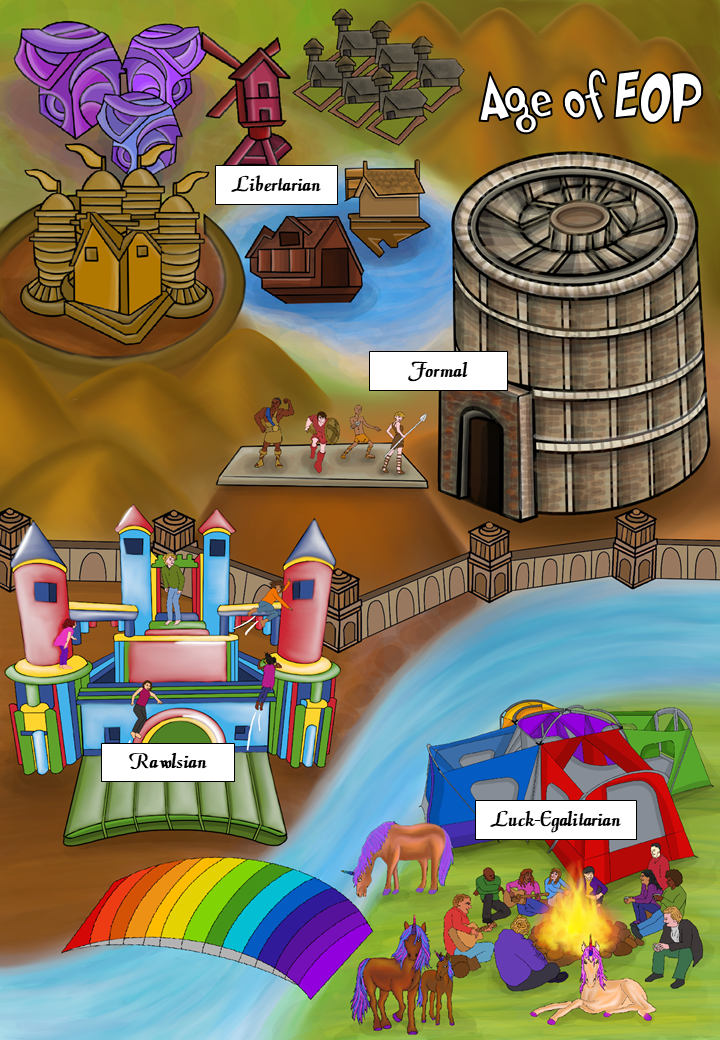}
\caption{A  depiction of popular EOP doctrines as settlements in an EOP empire \cite{fairfriends}.}
\label{fig:eop}
\end{figure}

Equality of Opportunity (EOP) is a helpful frame for understanding fairness concerns.  However, fair-ML literature has not yet established philosophically grounded and intuitively compelling connections between different EOP views and different fairness concerns. In this section, we introduce versions of EOP that are already influential in fair-ML and clarify the philosophical views that ground them. We will use the ‘EOP Empire’ \cite{fairfriends} shown in Figure \ref{fig:eop} to provide an accessible explanation of these doctrines.\par 

\subsection{Libertarianism}
Following~\citet{Arneson2018FourCO}, libertarianism has been introduced as a possible version of EOP.
The libertarian view emphasizes individuals' freedom to conduct themselves as they will, provided they do not violate others' basic rights against violence, theft, and other personal violations. For example, an employer is free to make a hiring decision that discriminates on the basis of race, or decline to make hiring decisions on the basis of job-relevant qualifications.  The emphasis is on people’s freedom, as is depicted in the unique architectural designs of the settlements of the libertarian village in Figure \ref{fig:eop}. Any holding of land acquired honestly (without theft or cheating) is claimed \emph{fairly}, even if it means that some end up living lavishly in ivory towers, while generation after generation of others live in dilapidated huts.\par 
A simple conceptualization of the libertarian view is a game of Monopoly — players are free to capitalize on whatever opportunities they have access to, such as rolling doubles and getting to move twice or picking up that Chance card that advances you to Boardwalk — provided they gain such access fair and square — no cheating by rolling biased dice, stealing from the bank or coercing players into making trades. All players are free to decide which property to chase. Whether they actually get the opportunity to buy and develop on that spot is not entirely devoid of chance, but the game does not attempt to correct for it. Instead, the emphasis is on players’ freedom to exercise their skills of negotiation and dice-throwing.\par 
While the libertarian view been interpreted as a fairness-enhancing perspective in fair-ML \cite{heidari_EOP}, this is not a form of EOP at all — there is nothing being equalized, nor does libertarianism satisfy EOP's characteristic commitment to eliminating irrelevant and arbitrary barriers to achievement. The libertarian view only focuses on ensuring a limited notion of procedural fairness— it would object to illegal or unfair means of gaining access to opportunities.

\subsection{Formal EOP}
Formal EOP emphasizes the distribution of desirable positions according to individuals’ relevant qualifications — in any contest, the most qualified person wins. This view rejects any qualifications that are irrelevant for the job. For example, hereditary privileges or social status, like being an aristocrat, won’t get you the job. In a formally \emph{fair} society, individuals are evaluated for positions based on qualifications relevant to those positions only. However, formal EOP makes no attempt to correct for arbitrary privileges or unequal access to opportunities that can lead to disparities between people’s qualifications. For example, because of the way that society has been set up, being a woman might preclude you from getting certain opportunities to build qualifications. The resulting disparities in your qualifications compared to male candidates is not something the formal view seeks to correct for. Instead, it emphasizes that the decision to give you the job will not be explicitly based on your gender (an irrelevant characteristic in the context of hiring). \par
A clarifying conceptualization of formal EOP is the Colosseum shown in Figure \ref{fig:eop}, which has its doors open to all talents. Everyone willing and able is welcome to  compete on the basis of their physical strength, but you compete with what you have — no special treatment once you are in. This means that even a pauper off the street is allowed to sign up, but whether they can actually put up a fight — without armor or weapons of any kind— against warriors brought up in the gladiatorial arts since childhood \cite{williams_1973}, is not something formal EOP is concerned with.  

In comparison to systems of hereditary privilege and entrenched discrimination, the idea of a society where ``people get ahead by being good at things'' amounts to remarkable moral progress \cite{markovits}. Formal EOP moves us toward decision-making processes based on relevant criteria and away from blatant discrimination.  As such, formal EOP—the rejection of baseless privilege in favor of qualifications-based advancement—arguably deserves a place as a baseline for fairness in machine learning.

\subsubsection{Fairness through blindness}
Formal EOP is easy to dismiss as a too-weak interpretation of EOP.  In fair-ML, formal EOP has been codified as \emph{fairness through blindness} \cite{dwork_awareness} — any protected attributes such as gender or race or disability status are deemed \emph{irrelevant} and are stripped away from the data. While it may be impossible to build genuinely gender/race/disability-blind models — and we have plenty of experiments showing that classifiers are capable of reconstructing protected attributes from redundantly encoded features \cite{dwork_awareness, Lipton_DLP} — \emph{fairness through blindness} can still pack a punch by explicitly recommending a list of irrelevant characteristics to exclude from consideration. To the extent that irrelevant characteristics (and their proxies) can be successfully excluded from an algorithm’s pipeline, formal EOP can make progress toward its aim of rejecting the use of socially important markers like race, gender, and social class as the basis for awarding privileged positions. 

For example, take the U.S. ``Ban the Box” campaign, that aims to pass laws requiring employers to be \emph{blind} to candidates’ criminal histories during initial assessments of qualifications.\footnote{\url{https://bantheboxcampaign.org}} The campaign aims to eliminate the check box on job applications that asks applicants to indicate whether they have a criminal history. Excluding criminal history from initial screenings of candidates is a powerful fairness-enhancing intervention that ensures that justice-involved people are not dismissed out of hand during the hiring process, allowing an applicant to be judged on the basis of their qualifications.\footnote{\url{https://www1.nyc.gov/site/cchr/media/fair-chance-act-campaign.page}}

Credit checks operate similarly to criminal history checks to exclude job candidates with poor credit scores from consideration. Where legal, credit checks allow employers to casually exclude candidates with imperfect credit from jobs for which they may be well qualified.  Where illegal, such as in New York City, employers are legally required to be \emph{blind} to job candidates’ credit histories during the hiring process.\footnote{\url{ https://www1.nyc.gov/site/cchr/media/credit-check-law-for-employees.page}}  Building algorithms that comply with laws such as New York City’s \emph{Fair Chance Act} and its \emph{Stop Credit Discrimination In Employment Act} would seem to require hiring algorithms to be blind to criminal and credit histories. Adding ``credit-related information'' to the set of irrelevant characteristics that are not used to evaluate candidates while hiring is another example of formal EOP at play.

\subsubsection{Test validity}
Formal EOP aims to consider only relevant characteristics for a position and to select candidates accordingly.  This aim lends objectivity to evaluations and excludes capricious or random decision-making. In addition to \emph{fairness through blindness}, formal EOP appears to demand that we also measure tests’ validity for members of different demographic groups, to ensure that a test is attentive to only relevant characteristics for members of all groups. 

Suppose a test systematically underestimates individuals’ qualifications in a way that tracks race or socio-economic status. For example, think of the SAT as a predictor of college success: when students can afford to do a lot of preparation, scores are an inflated reflection of students’ college potential.  When students don’t have access to preparatory material, the SAT underestimates students’ college potential. The SAT systematically overestimates more privileged students, while systematically underestimating less privileged students. \cite{Fishkin2014Bottlenecks} The test’s validity as a predictor of college potential varies across groups, and this would be a violation of formal EOP— Applicants should only be judged by ``college-relevant" qualifications, but the SAT’s accuracy as a yardstick for college potential varies with students’ irrelevant privilege.
Formal EOP instructs us to track whether a selection process is equally accurate for members of different social groups.  A test whose results are systematically skewed in a way that tracks irrelevant protected characteristics is not formal EOP compliant, even if the test is blind to group membership.

\subsubsection{Limitations of formal EOP}
Formal EOP has two key shortcomings— let’s call them the ``before'' and ``after'' problems. To motivate the “before” problem, recall the basic promise of formal EOP: it aims to reward individuals for their relevant skills and ends the aristocratic practice of including and excluding people from desirable positions based on family status or other irrelevant privileges. Before competitions, people are allowed to exercise their privilege to develop relevant qualifications. Take Bernard Williams’s famous example of a warrior society \cite{williams_1973}: formal EOP is completely satisfied in a society where any child can compete to be a warrior, even if only children of warriors get enough food and training to develop the combat skills necessary to win competitions.  The battle may be formally \emph{fair}, but the deck is stacked in favor of the children of warriors. Arbitrary and morally irrelevant privileges weigh heavily on the outcomes of formally \emph{fair} competitions because people can leverage them to build qualifications in advance of competitions. This undermines the promise of formal EOP —that only relevant skill will be rewarded— because it fails to stop gains from being distributed along the lines of privilege and disprivilege. 

Formal EOP also has an ``after'' competition problem.  After formally \emph{fair} competitions, winners are set up for even more success.  A candidate that is hired for a job is consequently granted access to more training and job experience. This makes them even more competitive in the next competition for jobs. The snowball effect also makes losers lose faster.  Suppose a formally \emph{fair} algorithm evaluates a candidate for a loan, and seeing their bad credit, grants a loan at a usurious rate— say 5 million percent interest! This makes it hard to pay back the loan, and easier to default. Defaulting makes their credit score even worse.  So, the next time the formally \emph{fair} algorithm assesses the candidate, it will give them an even worse loan based on their worse credit score.  This is the snowball effect.  Formal EOP compounds both privilege and disprivilege.

 Let’s put together the ``before'' problem and the ``after'' problem. In our society, outright discrimination leads to some people enjoying opportunities that are denied to others. Say, for example, in the investment banking world, women are not offered the professional opportunities available to their male colleagues.  As a result, women bankers amass fewer competitive qualifications than their male peers. They have fewer responsibilities, smaller portfolios, and slower promotions. Now, if we introduce a formally \emph{fair} competition for a high-ranking banking job, due to the ``before'' problem, formal EOP lets candidates convert their privileges into qualifications in advance of the competition. So, women enter this competition, and they tend to lose, (formally) fair and square, to men who have converted their privilege into qualifications. Now add in the snowball effect— Over time, women fall further and further behind in formally \emph{fair} competitions, simply on the basis of their lesser qualifications.  Over time, gender-based differences in qualifications appear to explain the disparity between men and women’s outcomes in investment banking, and the initial discrimination against women recedes from view. 
 
Elizabeth Anderson calls this ``discrimination laundering''~\cite{10.2307/j.ctt7t225}. As she puts it, with respect to racial discrimination, ''discrimination in on-the-job training is illegal; however, discrimination on the basis of differences in human capital due to differences in on-the-job training is not''~\cite{10.2307/j.ctt7t225}. The key takeaway is that formal EOP cannot fix these problems---its strengths are limited to measuring people’s actual qualifications accurately, and in excluding irrelevant information.  The problem we have just put our finger on is the inequality in \emph{developmental opportunities}---women, racial minorities, and other marginalized groups are historically discriminated against and have lesser opportunities to develop qualifications. To address the problem of unequal access to developmental opportunities---especially for historically disadvantaged and marginalized demographics---we need substantive EOP. 

\subsection{Substantive EOP}
Substantive EOP aims to provide people genuine opportunities to build their qualifications, to give them  a \emph{fair} chance of success  at the point of competition. We will consider two versions of substantive EOP that are particularly influential in fair-ML: Rawls's fair EOP and luck-egalitarian EOP.

\subsubsection{Rawls's fair EOP}
Rawls’s fair EOP~\cite{Rawls1971Justice} maintains that all people, regardless of how rich or poor they are born, should have opportunities to develop their talents, so that people with the same talents and motivation have the same educational and employment opportunities. His moral philosophy can be summarized as ``Equally talented babies should have equal life prospects''. Rawls wants to ensure that your privileged birth doesn’t snowball into a lifetime of privileges that allows you to outcompete children whose disadvantage at birth has led to compounded disadvantages.  

One can picture the Rawlsian settlement in an EOP empire as a bouncy castle, shown in Figure \ref{fig:eop}. The bouncy castle of social security has strategically placed trampolines to ensure that, no matter people’s starting points in life, children with the same talents and willingness to use them have the same opportunities to propel themselves to success. This emphasizes the fact that Rawls’s view is targeted at the equality of developmental opportunities —access to opportunities to develop qualifications from childhood onward.  

\begin{figure}[ht]
\centering
\includegraphics[scale=0.073]{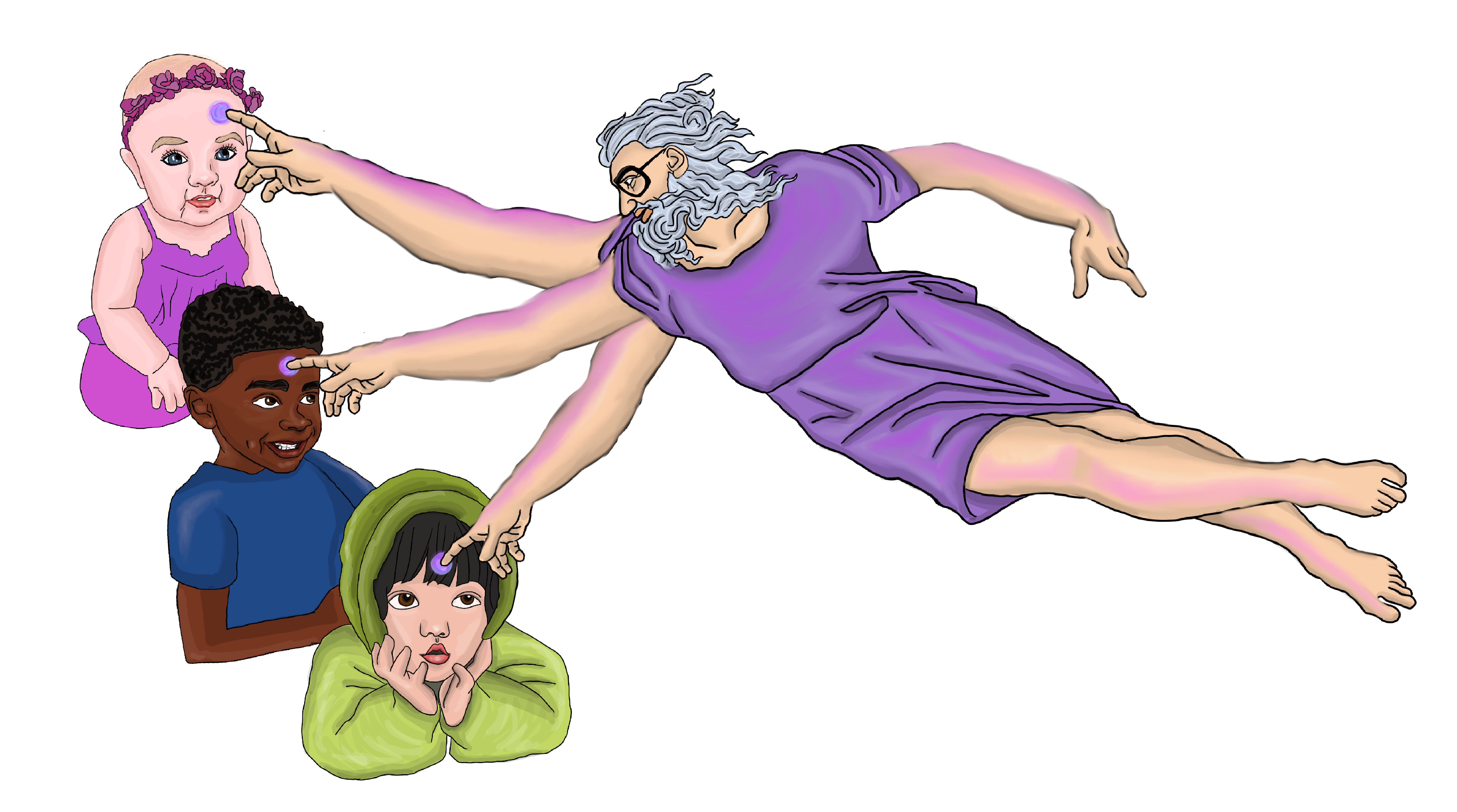}
\caption{'Rawls and Adam' depicts John Rawls peering into the mind of three children, to understand their native talents and ambitions ~\cite{fairfriends}.}
\label{fig:rawls}
\end{figure}

Another clarifying conceptualization is shown in Figure \ref{fig:rawls}, where John Rawls attempts to peer into the mind of a child. Rawls’s fair EOP emphasizes that people should be given access to the opportunities they desire so that they can build up their qualifications. In Figure \ref{fig:rawls}, we see Rawls attempting to understand the native talents and ambitions of a child, so as to provide them with all the developmental opportunities they desire, to hone their talents. 

Rawls’s fair EOP regulates individuals' access to developmental opportunities from childhood onward. However, at the point where an ADS is making a decision, it is already too late to provide people with opportunities to build qualifications. Instead,  fair-ML has reinterpreted Rawls's view as requiring that a competition for positions must measure candidates on the basis of their talents and motivation, while ignoring qualifications that reflect candidates' unequal developmental opportunities prior to the point of competition. Along these lines, popular fairness notions that codify Rawls’s fair EOP include statistical parity \cite{dwork_awareness} and equality of odds \cite{hardt_EOP2016}. These formulations assume that talents and motivation are equally distributed among sub-populations. Under this assumption, if competitions are adjudicated on the basis of talents and motivation, each sub-population should have the same success rate as any other. \par

\subsubsection{Luck egalitarian EOP}
Luck egalitarians maintain that Rawls does not go far enough in controlling for factors that provide unfair advantage or disadvantage. This view, due to \citet{dworkin_1981}, maintains  that a  person's outcomes should be affected only by effort or ``choice luck'' (responsible choices).  No effects of circumstance, or ``brute luck'' --- from having rich parents to getting struck by lightning— should influence the outcome of a \emph{fair} competition. 

An accessible explanation of the luck egalitarian view, shown in Figure \ref{fig:eop}, is an open campsite. Settlers gather around a communal fire, having forsaken all disparities in wellbeing due to unchosen circumstances, in favor of unicorns on rainbows, 
and outcomes that reflect their effort. 

The luck egalitarian doctrine is extremely attractive. But the hard question now is— can we separate the effects of brute luck from the effects of responsible choices? \citet{Roemer2002} gives us a version of EOP that can be operationalized, and has been codified in fair-ML by \citet{heidari_EOP}. Roemer recognizes that a person’s effort is affected by their circumstances. Roemer’s EOP doctrine asserts that instead of trying to separate a candidate’s qualifications into effects of brute luck and effects of choice luck, we can control for certain matters of brute luck —such as race, gender and disability— that we know impact a person’s access to opportunities to develop qualifications. We can conduct a \emph{fair} competition, by ranking candidates in their effort-circumstance distribution— comparing people with others of the same circumstances.

Roemer’s EOP bypasses the non-intuitive idea that we can separate a person’s qualifications into a responsible effort component and an irrelevant circumstances component. Instead, he compares people only to others in similar circumstances, effectively controlling for irrelevant circumstances— apples are compared to apples, oranges to oranges and bananas to bananas. 

Notably, Roemer’s view comes with a cost: it creates an apples and oranges problem. How do we compare and select among the top-scoring apples and the top-scoring oranges? Substantive EOP doctrines do not give us any guidance on this front, and so they fall short of providing a complete ranking of all candidates.

\paragraph{Separating the effects of effort and circumstance}
Roemer’s view is intuitively appealing because it declines to split a person’s qualifications into  responsible effort and irrelevant circumstances. This is important, for two reasons.

First, effort and circumstances interact: responsible effort and arbitrary circumstances are impossible to disentangle.  Take the example of a child who grows up in a circus family. Family circumstances are a  matter of brute luck.  In this  case, these circumstances supply opportunities for the child to learn the family  business. However, suppose as well that the child also has a talent for acrobatics, and is motivated to practice — they put in the responsible effort. Seeing this, the family nurtures the child’s gift (lucky circumstance).   In response, the child increases their efforts (responsible effort). Is the child’s success as an acrobat attributable to circumstance --- being born into a circus family? Yes. Is the child's success the fruit of their own labor? Also, yes. It takes responsible effort to excel in acrobatics. We simply cannot separate the child’s qualifications that are due to responsible effort from those that are due to brute-luck circumstances. 

Moreover, we argue that it is not inherently problematic that acrobat families raise acrobat kids, or that farming families pass on the farming trade. Decoupling the effects of effort and circumstance is only clearly necessary when circumstances track access to a broad range of desirable positions in society. If a person's family background, or other matters of circumstance, set them up to surpass \emph{all} other candidates in almost \emph{every} arena, we have a \emph{fairness} problem: it indicates that circumstances (arbitrary  privilege), and not the relevant skills, are the basis for deciding competitions. 

Conversely, if a person's race impedes their ability to compete in \emph{any} arena, this suggests that their relevant skills are not being taken into consideration: arbitrary circumstances are ruling out a wide range of life paths, irrespective of responsible effort and the qualifications it has produced. When circumstances of birth rule out a wide range of desirable positions, we also have a problem of systemic \emph{unfairness}. By contrast, the limited amplification of effort by circumstances---the acrobat who gets a leg up from their circus family--- may or may not indicate the existence of \emph{unfairness} that a society must counteract as a matter of justice.  

The second reason to shift substantive EOP away from splitting a person’s qualifications into responsible effort and  brute-luck circumstances has to do with the needs of decision-makers.  For example, from an employer’s perspective, a good hiring decision is contingent on evaluating all of a person’s job-relevant qualifications, not only those qualifications for which they are responsible. If we insist that substantive EOP pay attention to only a part of a person’s qualifications, this puts \emph{fair} decision-making on a collision course with prudent decision-making. To run functioning workplaces, employers need candidates who \emph{are} qualified, not candidates who \emph{would have been} qualified under different circumstances. To stay  afloat, banks need to extend loans to customers who will repay them, not to customers who would repay their loans were their unlucky circumstances different. The further substantive EOP’s counterfactual assessment of a person’s qualifications diverges from a person’s actual qualifications, the less value those assessments have to employers, lenders, and other decision-makers.\par

\section{The EOP Spectrum}
\label{sec:spectrum}
\begin{figure}[h]
\includegraphics[scale=0.34]{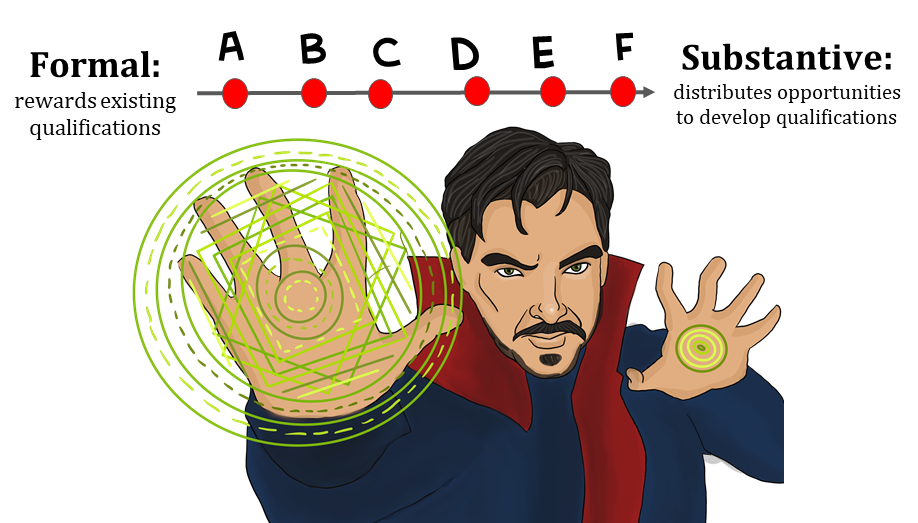}
\caption{The EOP spectrum ~\cite{fairfriends}.}
\label{fig:eop_spectrum}
\end{figure}

Using the taxonomy of fairness ideals that we have introduced so far, we now introduce a framework to guide the selection of a suitable fairness measure for a given context. To do this, we first introduce the idea of an \emph{EOP spectrum}, shown in Figure \ref{fig:eop_spectrum}. On one end, we place the doctrine of formal EOP, which rewards candidates on the basis of their \emph{existing (relevant) qualifications}. Rejecting formulations of substantive EOP that split a person's qualifications into relevant effort and irrelevant circumstances pieces, we interpret substantive EOP in the spirit in which it was proposed: as a principle that distributes developmental opportunities, and place it on the other end of the EOP spectrum. In particular, we view substantive EOP as the principle that distributes opportunities to candidates to develop qualifications \emph{in advance of future competitions}.

We argue that practical fairness concerns will lie on this EOP spectrum, and by identifying which outcome we find more morally suitable in a particular context—rewarding past qualifications or providing qualifications-building opportunities to promising candidates—we can select a suitable fairness ideal.

We evoke the image of the fictional Sorcerer Supreme, shown in Figure \ref{fig:eop_spectrum}, to intuitively explain how the EOP spectrum works. The Sorcerer Supreme has the ability to control time--- to see how a decision made now provides developmental opportunities in the \emph{future}, and how candidates’ qualifications now have been shaped by privilege or disadvantage in the \emph{past}. The choice he is faced with is whether to reap the benefits of skills already assembled, or to delay the benefits to a later time, focusing instead on fortifying power. This is exactly the choice a decision-maker, interested in making \emph{fair} allocations, is faced with. 

To the far left, we have the formal point of view. The emphasis is on accurately measuring the candidate’s relevant current qualifications. Social inequalities may have precluded candidates from certain demographics from building their qualifications, but this is not a \emph{fairness} concern from a formal EOP standpoint. We absolve ourselves of the responsibility to correct for such inequalities at this time, and instead focus on finding the most qualified candidate.  

To the far right, we have the substantive view, which treats every competition/selection process as an occasion to equitably distribute opportunities to develop skills and qualifications. Purely substantive EOP extends access to desirable positions in a manner that equalizes the distribution of developmental opportunities among the applicant pool---In a purely substantive world, every hiring, loan, and school admissions decision is an opportunity to invest equitably in people’s qualifications and life prospects.  

Clearly, neither of these extremes — purely formal or purely substantive EOP — are desirable. In between, we have decision-making geared partly at identifying and rewarding qualified candidates, and partly at ensuring that our society invests equitably in developing all citizens’ qualifications and life prospects. Using our framework, a decision-maker can choose to evaluate based on existing qualifications or to invest in building promising candidates’ skills-- or some combination of the two. 

For example, consider a hiring decision governed mostly, but not entirely, by substantive EOP. A law firm is looking for paralegals and it decides that, practically speaking, the minimum qualifications that they require from applicants is that they must be trained and certified paralegals. Adopting a substantive approach to \emph{fair} hiring, the firm shortlists all the candidates who meet this bar and then selects and interviews candidates by lottery — the firm is willing to train candidates who have the skills required to be certified. 
This is a decision process that recognizes that this hiring decision will affect people’s opportunities to develop skills and compete for positions later on. It’s also a decision that recognizes that the qualifications that people bring to this moment reflect their privilege or disadvantage. The law firm is being \emph{fair}, while also insisting on the qualifications it needs. And the law firm can make good hiring decisions: it does not have to ignore some undeserved part of candidates’ qualifications. As a decision-maker, we haven’t split candidates’ qualifications into a relevant effort piece and an irrelevant circumstance piece.  Instead, we have split the difference between a formal EOP competition and a substantive EOP distribution.

\section{Re-interpreting Impossibility Results} 
\label{sec:impossibility}
The impossibility results \cite{chouldechova_impossibility, Kleinberg_impossibility, friedler_impossibility} are commonly interpreted to mean that \emph{fairness is impossible}. But, if we view different fairness measures as promoting different conceptions of EOP (formal vs substantive), then this incompatibility is wholly unsurprising. From a philosophical perspective, the idea that values trade off against one another is expected.

Consider the EOP-based fairness views that we would expect to be in conflict.  Formal EOP rewards people’s relevant qualifications, while substantive EOP distributes developmental opportunities equitably. In fair-ML, formal EOP metrics focus on correctly measuring a person’s relevant qualifications. Codifications of substantive EOP, on the other hand, either adjust people’s qualifications to account for their undeserved circumstances or seek to equalize developmental opportunities at the group level. It is unsurprising that formal and substantive EOP measures would trade off against one another. These views pull in opposite directions, and their corresponding metrics cannot simultaneously be optimized.\par
We argue that this incompatibility does not impede the practitioner’s quest for \emph{fair} decision-making. Some circumstances call for a more formal approach, while others merit a substantive approach. When it comes to \emph{fairness}, we must make value-based judgements about which approach---or how much of each approach--- is suitable for a given context. This is where our framework, based on the EOP spectrum, can be helpful. 
\section{Normative Guidance to Select a Fairness Ideal}
\label{sec:examples}
The mutual incompatibility among different fairness ideals is neither surprising nor necessarily worrisome: it simply requires taking a stand about what matters from a moral standpoint. Grounding fairness ideals in EOP doctrines, and using the proposed EOP spectrum, shows us the intuitive appeal of different fairness ideals for different contexts, which we will now demonstrate using several hypothetical and real-world examples. Figure \ref{fig:eop_spectrum} shows where each of these scenarios fall on the EOP spectrum.

\subsection{Fair footraces (A)}
Consider the hypothetical situation in which a king wants to prove his mettle, by conducting a footrace in the kingdom. How do we model a \emph{fair} footrace? A purely formal approach might be sufficient: No bars to entry, both aristocrat and peasant is eligible to compete, even if it means that some are attended to by an army of servants, while others run bare-footed. We are not concerned about finding the most promising athlete and grooming him to be the leader of the king’s guard. All we want is to see whether someone can outrun the king, and, to this extent, the competition is \emph{fair}.

\subsection{Loan decisions (B) }
Let’s look at a real-world example now — making \emph{fair} loan decisions. As a bank, we care about an applicant’s credit-worthiness at the time of applying and the risk of them defaulting on the loan— it is not our primary concern to distribute opportunities to candidates to build up their credit-scores. However, we still want to be \emph{fair} in our allocation of loans, and here formal EOP as \emph{fairness through blindness} can pack a punch. We can mandate that decision-makers —whether algorithmic or human, or some combination of the two— must not take into account irrelevant characteristics, such as race, gender, disability status, religion, etc while making an assertion of credit-worthiness. Formal EOP as \emph{test validity} also enhances our fairness criteria in this scenario; Disparate impact in loan decisions --- where decisions are systematically skewed against applicants of a certain demographic --- would violate the principle of formal EOP, because this would mean that the test does not accurately measure the relevant qualifications of all applicants. 

\subsection{Blind auditions (C)}
A more compelling example of \emph{fair} decision-making is 'blind auditions’ in talent shows. Say, we are looking to find the fresh new face of pop music, and to propel them to the top of the charts. It is too late to spend years and years training candidates to expand their vocal abilities, and so we only concern ourselves with \emph{fairly} evaluating candidates today. To make sure that judges are not swayed by irrelevant traits like gender, race, and appearance, we conduct blind auditions, and evaluate contestants solely on the relevant rubric in this competition— their musical talents. The blind audition embodies the strength of formal EOP. But our judges also realize that being on a talent show is an important opportunity to practice performing in front of a live audience and to get mentorship and vocal training. While they select candidates on the basis of their vocal abilities, they also make a point of selecting candidates who will benefit from such development  opportunities. In doing so, judges also uphold substantive ideals of \emph{fairness}, such as candidates' equitable access to developmental opportunities.

\subsection{Hiring (D)}
Let’s look at hiring. Say, a company is looking to ramp up its digital footprint, for which it needs to hire a new engineering unit.  The company decides that, from a practical standpoint, all it needs in an employee is knowledge of the fundamentals of software engineering and the ability to code. The company is willing to hire, and train all the candidates who have the necessary coding skills, whether or not they have a relevant college degree. Adopting a substantive approach to \emph{fair} hiring, the company gives all applicants a short take-home assignment, and accepts all the candidates who successfully complete it. Based on how many candidates pass this initial screening, it shortlists and interviews candidates by lottery. Or it decides to shortlist a diverse set of candidates that will start to remedy the demographic imbalance in this office. This is a decision-making system that recognizes that this hiring decision will affect people’s opportunities to develop skills and compete for positions later on. It’s also a decision that recognizes that the qualifications that people bring to this moment reflect their privilege or disadvantage—not all applicants could afford going to university. The company is being \emph{fair}, while also insisting on basic relevant qualifications. And the company is able to find qualified candidates by this procedure— it does not have to ignore some undeserved part of candidates’ qualifications, or attempt to separate a candidate’s qualifications into a relevant effort piece and an irrelevant circumstance piece.

\subsection{College Admissions (E)}
Let’s look at the situation of college admissions. The aim of the admissions committee is to admit students that they think will excel in the program — on the basis of their potential for future success, not purely on their academic accomplishments up to that point. The substantive view is immediately attractive; Going to university is an important opportunity to build qualifications, and in making a \emph{fair} admissions decision the committee is deciding how to equitably distribute access to developmental opportunities.   
Unlike in the situation of hiring, where companies wholly bear the cost of up-skilling their employees, in college admissions, the cost of building future qualifications is borne jointly by the university and the student — the university puts its resources and reputation on the line for the student, while the student bears the monetary cost (and opportunity cost) of joining the program. This cost-sharing creates a strong incentive for the admissions committee to take a more substantive approach— they can prioritize finding a diverse cohort of qualified applicants. Roemer’s EOP gives us a simple procedure to do this: rank applicants among others with the same circumstance —such as percentile of test scores, ranking of high school, socioeconomic status, research experience, etc— and admit the top candidates of each group. By taking a substantive approach to \emph{fair} college admissions, we are able to admit a varied cohort of qualified students.

\subsection{Saving Sisyphus (F) }
We evoke the popular Greek myth of Sisyphus— who is condemned to roll a boulder up a mountain, only to have it roll back down each time he reaches the top—to motivate a situation where a purely substantive approach might be necessary. In a society with deep-rooted, structural discrimination, entire swaths of people might be excluded from developing competitive qualifications, purely due to circumstances of birth. If candidates from a certain demographic group are systematically ruled out in competitions for a wide range of desirable problems, then we have an \emph{unfairness} problem. As stated earlier, relevant skills are not being used as the basis for decision-making. To break the Sisyphean struggle of candidates from marginalized groups in society, we might want to employ a purely substantive approach, where competitions are used to equitably distribute developmental (qualifications-building) opportunities.   
\section{Beyond Fairness: Broader Justice Considerations}
\label{sec:justice}

Taking a broader view of EOP doctrines brings to the forefront questions about social justice, and its relation to \emph{fairness} in automated decisions systems. Fairness concerns are a subset of justice concerns. To ensure that that algorithms are broadly just, and not only \emph{fair}, we need to think about the other requirement of justice that might apply to algorithms. In his theory of justice, for example, Rawls adopts not only the EOP principle, but two other principles that must be satisfied for a society to satisfy standards of justice for democratic societies.  He arrives at these principles, which together govern the distribution of rights, opportunities, income, and other socially produced goods via the original position ~\cite{Rawls1971Justice}. 

\begin{figure}[h]
\centering
\includegraphics[scale=0.075]{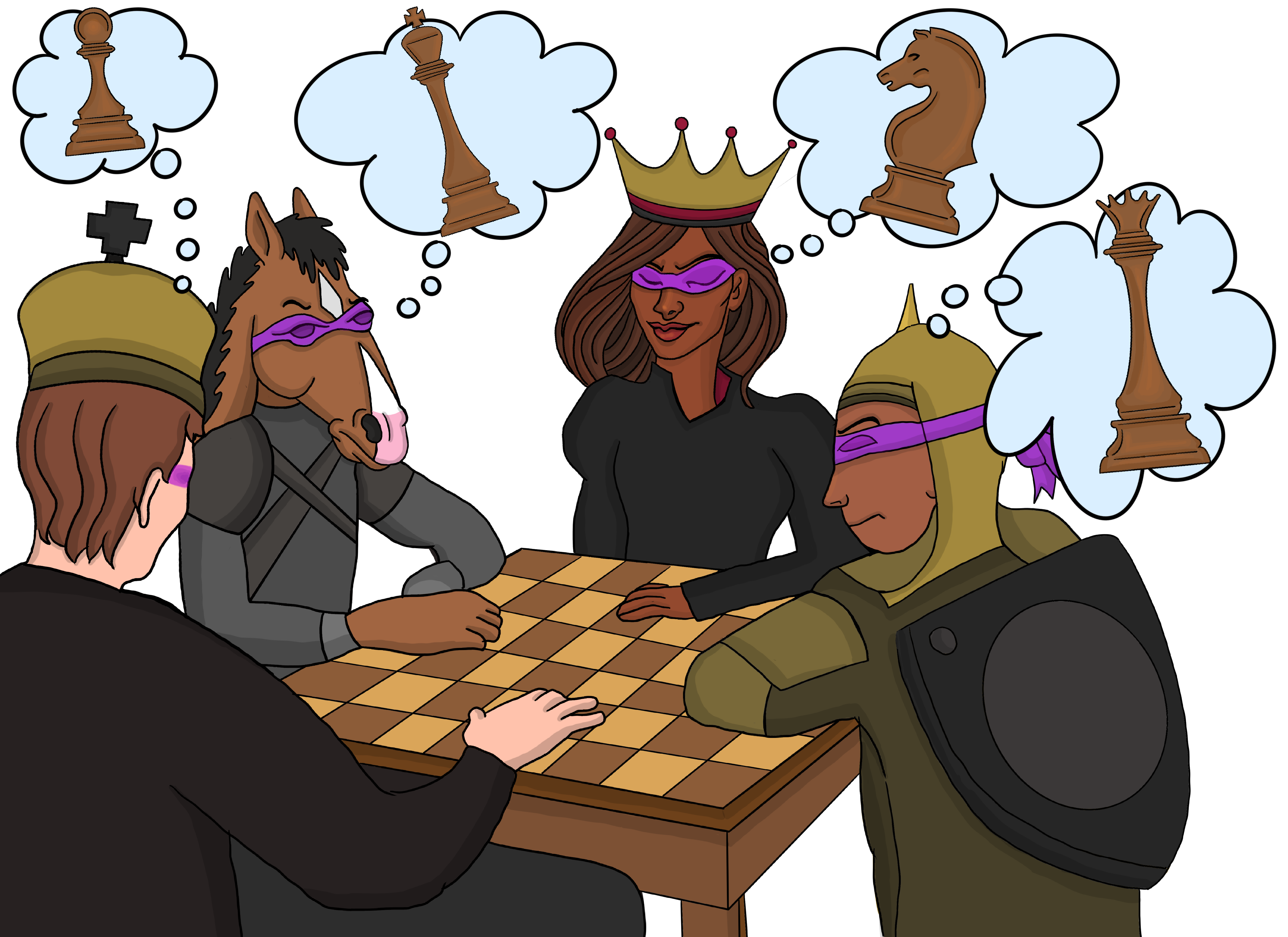}
\caption{Rawls's Veil of Ignorance, depicted as a negotiation between different pieces in a game of chess ~\cite{fairfriends}.}
\label{fig:veil}
\end{figure}

\subsubsection{The original position}
The original position is a thought experiment intended to illustrate how the free and equal citizens of a democracy might come to agreement about principles of justice to govern their society. The key feature of the original position is the \emph{veil of ignorance}.  The veil deprives participants in the original position thought experiment of knowledge of their gender, talents and other personal characteristics, as well as knowledge of their circumstances (e.g., family background, socioeconomic status) to ensure that participants can think impartially about the principles that will govern a \emph{just} society---from the operation of courts to the tax code in their society.

Participants in the original position thought experiment, like pieces on a game board, shown in Figure \ref{fig:veil}, know the rules of the game and as individuals, each has a personal conception of what \emph{success} looks like. However, they do not know which piece they will end up being, and must negotiate how to arrange the powers of movement and gameplay attached to different pieces (positions in society) through the \emph{veil of ignorance}. Rawls posits that the principles of social cooperation that result from this negotiation will protect the interests of all citizens. 

\subsubsection{Rawls's theory of justice}
Rawls proposes the following principles in his theory of justice \cite{Rawls1971Justice} as the (or one possible) outcome of the original position:

\begin{enumerate} 
    \item Rights and liberties: Everyone has the same inalienable right to equal basic liberties.
    \item
    \begin{enumerate}
        \item Rawls's fair EOP: All offices and positions must be open to all under conditions of fair equality of opportunity.
        \item Difference principle: Any social inequality must be applied in such a manner that they be of the greatest benefit to the least advantaged.
    \end{enumerate}
\end{enumerate}

 In the Rawlsian system, these principles are hierarchically ordered – fair EOP has less priority than the principle of basic rights and liberties. This means that we cannot go about creating fair EOP in such a manner that would encroach on a fundamental right or liberty.
 
 In the context of algorithms, this broader perspective is helpful to see how an ADS that is (statistically) \emph{fair} can go on to infringe on basic rights and liberties and, in effect, be \emph{unjust}. Take the example of \emph{fair} hiring of people with disabilities. “Disability” is legally protected characteristic, and a philosophically irrelevant matter of brute luck/circumstance, and would be removed from explicit consideration by a \emph{fair} ADS. Disability could be redundantly encoded in other information that is taken into account.
 
 For example, if social media information is used at the time of conducting background checks, the hiring algorithm could infer disability status based on membership in certain social groups or on posting about disability-related issues. Now, this disability-blind ADS could still discriminate on the basis of \emph{inferred} disability, which would incentivize people against joining such groups and speaking about such topics. Such an ADS could satisfy some conception of \emph{fairness} as EOP and yet be fundamentally \emph{unjust}: it would violate an applicant’s freedom of speech and freedom of association.
\section{Conclusion}
\label{sec:conclusion}

Applying a fairness-enhancing intervention requires us to take into account the incentives of the decision maker and the underlying socio-political dynamics at play. Doctrines from political philosophy are a useful source of normative guidance for which fairness ideal is suitable in which setting. In this paper, we first introduced a taxonomy of fairness ideals using doctrines of EOP from political philosophy, clarifying their conceptions in philosophy, and then proposed codification in fair-ML. We  arranged these fairness ideals onto an EOP spectrum, which serves as a useful frame to guide the design of a \emph{fair} ADS in a given context. We demonstrated the utility of our EOP-framework using several real-world and hypothetical examples. 

 EOP doctrines can offer us useful guidance in the design of \emph{fair} algorithmic systems, but they also come with limitations. Overlooking these limitations can embolden their application in spheres in which theory provides little to no guidance. Importantly, EOP doctrines do not give us any direction about \emph{where} to apply fairness criteria --- in the procedure or in the outcomes.  The guidance is only about \emph{how} a fair test should behave. Further, although we have presented examples where Roemer’s luck-egalitarian EOP is useful, trying to adjust a person’s qualifications for matters of brute luck/circumstance can lead us down the problem of comparing apples with oranges. We might be able to find the highest-ranking apples and the highest-ranking oranges using EOP doctrines, but we get no guidance about how to go about comparing the two.  
\balance

\bibliography{main}

\end{document}